\documentclass[epj]{svjour}

\usepackage{graphicx}
\usepackage{fancyhdr}
\def\makeheadbox{{
 }}
\def\mytitle{My title} 
\def\myauthors{My name}  
\def\mytype{My type of session}
\def\mysession{My session}

\def\mytitle{Search for Higgs decay to $\tau\tau$ at the Tevatron} 
\def\myauthors{Ilya Kravchenko}    
\def\mytype{Contributed Talk}    
\def\mysession{Colliders - Higgs Phenomenology}

\begin{document}
\title{Search for Higgs decays to tau lepton pairs at the Tevatron}
\author{Ilya Kravchenko for the CDF and D0 Collaborations
}                     
%
%
\institute{Massachusetts Institute of Technology
}
%
\date{}
\abstract{ We present a search for neutral supersymmetric Higgs bosons
  decaying to $\tau^+\tau^-$ pairs produced in $p\bar{p}$ collisions at
  $\sqrt{s}=1.96$~TeV. The data have been collected with the CDF II
  and D0 detectors at the Tevatron collider at Fermilab ($1$~fb$^{-1}$
  of integrated luminosity per experiment). No significant excess above
  the standard model backgrounds is observed. We set exclusion limits
  on the Higgs production cross-section times the branching fraction
  of its decay to $\tau^+\tau^-$ pairs for Higgs masses in the range from
  90 to 250~GeV/$c^{2}$. We also set exclusion limits on MSSM
  parameters $m_A$ and $\tan\beta$ in several benchmark scenarios.
} 
\maketitle
\section{Introduction}
\label{intro}

 The Higgs mechanism provides the scheme for the electroweak symmetry
breaking, that is accepted in the standard model (SM) as well as some
of its supersymmetric extensions, such as Minimal Supersymmetric
Standard Model (MSSM) \cite{mssm}. The Higgs boson, if it exists,
remains to be found, and the Tevatron experiments are actively
searching for it. Due to low production cross-section predicted for
Higgs in the SM, the SM Higgs cannot be observed with the amount of
data collected to date. The production of MSSM Higgs may be enhanced
relative to that of the SM by several orders of magnitude for certain
areas of MSSM parameter space \cite{mssm-xsec}, making discovery of a
MSSM Higgs at the Tevatron possible with less integrated luminosity.

  In MSSM, the Higgs sector is comprised of two doublets of complex
scalar fields. These correspond to the five physical particles: two
charged bosons $H^\pm$, one neutral CP-odd boson $A$ and two more
neutral CP-even $h$ (light) and $H$ (heavy) particles.  At the tree
level, two free parameters are sufficient to describe properties of
these bosons.  Most often, these parameters are chosen to be the mass
of the CP-odd state $m_A$, and the ratio of vacuum expectation values
of Higgs coupling to down-type and to up-type fermions denoted
$\tan\beta$. At low $m_A$ and high $\tan\beta$ the $A$ is almost
mass-degenerate with either $h$ or $H$, and the production
cross-section of the $\phi$ is enhanced by $\tan^2\beta$.  It is
conventional to use $\phi$ to denote any of the $A$, $h$ or $H$.

 The MSSM Higgs at the Tevatron is expected to be primarily produced
in the gluon fusion process, and the second most significant process
is $b\bar{b}\to \phi$. It decays roughly 90\%(10\%) of times into
$b\bar{b}$($\tau\tau$) pair. The analyses presented in this paper are
devoted to the latter channel.
  CDF has searched for the Higgs decaying into the
final states $\tau_e\tau_{had}$, $\tau_\mu\tau_{had}$ and
$\tau_e\tau_\mu$, where the subscript denotes either the corresponding
leptonic, or a hadronic $\tau$ decay. The D0 analysis uses the
$\tau_\mu\tau_{had}$ final state.

\section{Data and Monte Carlo Samples}

The analyses are based on the data collected at the Tevatron in
$p\bar{p}$ collisions at $\sqrt{s}=1.96$~TeV. Both CDF and D0
experiments used data samples that correspond to the integrated
luminosity of $\sim$1~fb$^{-1}$.

The data of CDF have been collected using di-lepton triggers for the
$\tau_e\tau_\mu$ final state, and lepton$+$track triggers. At D0,
single muon triggers have been used, that require hits in the muon
system in conjunction with a high-momentum track in the central
tracking volume.

  For background estimates and efficiency calculation both experiments
have used Monte Carlo samples created with PYTHIA event generator
\cite{pythia} and GEANT detector simulation \cite{geant}.

\section{Backgrounds}

The largest and irreducible background in this search comes from
events with $Z/\gamma^*\to\tau\tau$. This background is
estimated using Monte Carlo simulation.

 The QCD multi-jet events and $W+$jet events constitute the second
largest source of the background. These contributions can be
controlled with well chosen selection criteria.

Finally, there is a number of small backgrounds that are taken into
account, such as di-boson and $t\bar{t}$ production, and $Z\to
ee$,~$\mu\mu$.

\section{Analysis}

\subsection{Event Reconstruction and Selection}

  The analysis of each experiment is only briefly described in this
paper, a complete report can be found in \cite{CDF-ana} and
\cite{D0-ana} for the CDF and D0, respectively.

For the reconstruction of a leptonic tau decay trigger leptons are
used both at CDF and D0. These leptons are required to match a
high-momentum track. They also have to pass isolation requirements on
the total amount of energy deposited in the calorimeter and scalar sum
of charged track momenta detected by the tracking system in a cone
around the leptons.

The $\tau$ leptons decaying hadronically are identified by detection
of collimated jets. The primary $\tau$ decay channels contributing to
this signature are the $\pi^\pm\nu_\tau$, $\pi^\pm\pi^0\nu_\tau$ and
$\pi^\pm\pi^\mp\pi^\pm(n\pi^0)\nu_\tau$. The jets detected in
calorimeter are thus matched to one or three charged tracks, the
angular size of the jets is limited to a narrow cone and, in case of
CDF, isolation requirements on detected particles outside of the
$\tau$ jet cone are imposed.  The DO experiment identifies $\tau$
leptons with neural networks developed on $Z\to\tau\tau$ events.  With
such selection, multi-jet QCD backgrounds are suppressed.

Other backgrounds are reduced by several requirements at the event
level.  Events at D0 have to pass $W$-veto that removes $W\to\mu\nu$
events, with $m_W$ reconstructed using the momentum of the muon and
the vector of the missing energy. Similar backgrounds are suppressed
by CDF with the requirement that rejects events where the missing
energy points away from the combined direction of the reconstructed
$\tau\tau$ decay products. Other methods used by CDF and D0 to
eliminate lesser backgrounds can be found in \cite{CDF-ana} and
\cite{D0-ana}.

After applying event reconstruction and selection to the data samples,
both CDF and D0 find that the observed yields in the data match the
expected counts of background events within uncertainties 
(see Table \ref{tab:yields}). Note that CDF errors are
statistical only, while the errors of D0 also include systematics.

\begin{table}
\caption{Event count in data and predicted background}
\label{tab:yields}       
\begin{tabular}{lcccc}
\hline\noalign{\smallskip}
  & \multicolumn{3}{c}{CDF} & D0\\
  & $\tau_e\tau_{had}$ & $\tau_\mu\tau_{had}$ &  $\tau_e\tau_\mu$ & $\tau_\mu\tau_{had}$ \\
\noalign{\smallskip}\hline\noalign{\smallskip}
Backg.    & 1196$\pm$19 & 1002$\pm$13 & 369$\pm$4 & 1287$\pm$130\\
Data      & 1215        & 1000        & 374       & 1144 \\
\noalign{\smallskip}\hline
\end{tabular}
\end{table}

\subsection{Results for  $\mathbf{\sigma(p\bar{p}\to \phi X)\cdot Br(\phi\to\tau\tau)}$}

The selected data samples have been searched for the signs of the
Higgs signal in the mass range from 90 to 250(200)~GeV/$c^2$ by
CDF(D0).


\begin{figure}
\includegraphics[width=0.45\textwidth,angle=0]{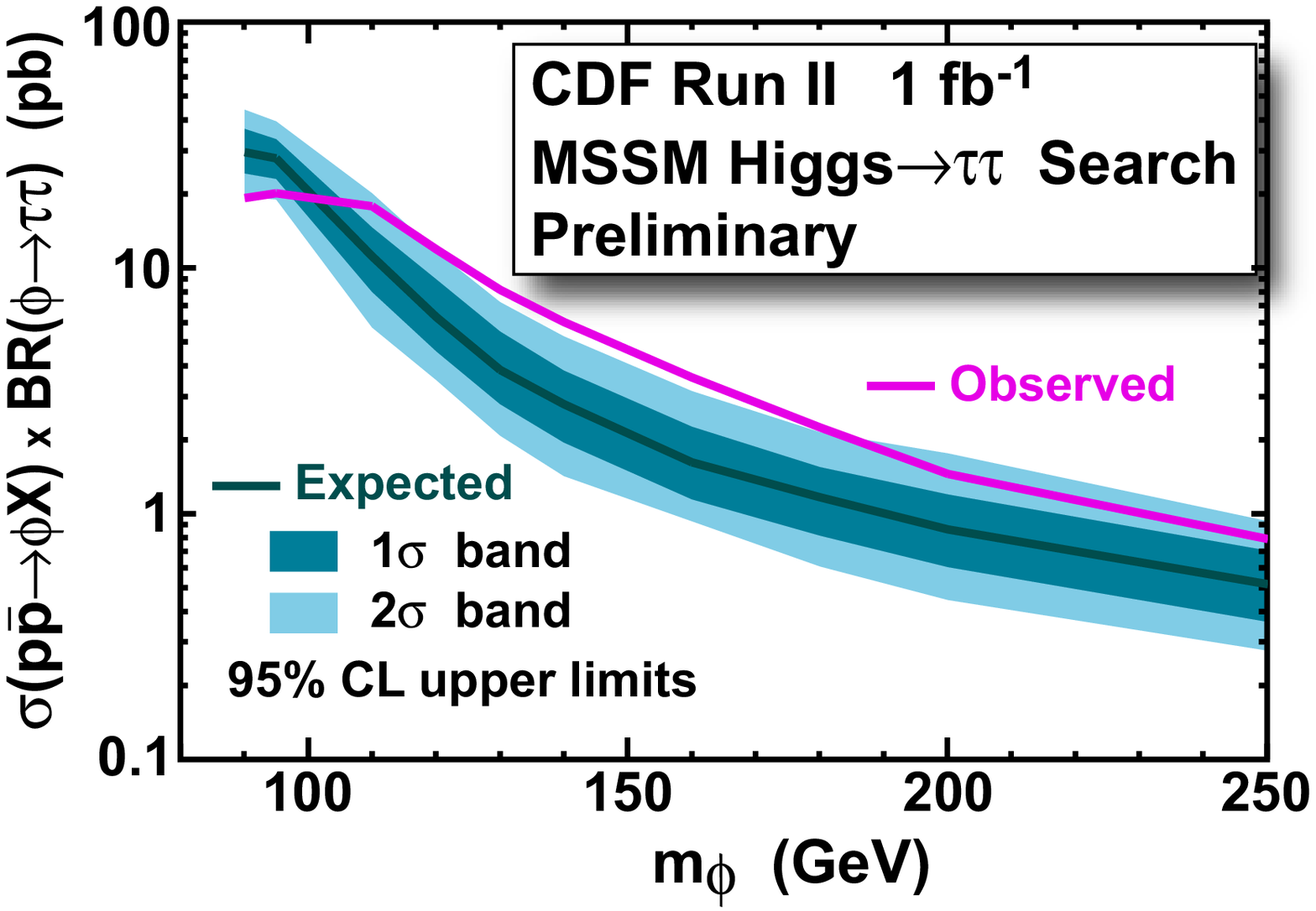}
\caption{The  limit on $\sigma(p\bar{p}\to \phi X)\cdot Br(\phi\to\tau\tau)$ from CDF
as a function of Higgs boson mass.
}
\label{fig:CDF-xsec}       
\end{figure}

\begin{figure}
\includegraphics[width=0.45\textwidth,angle=0]{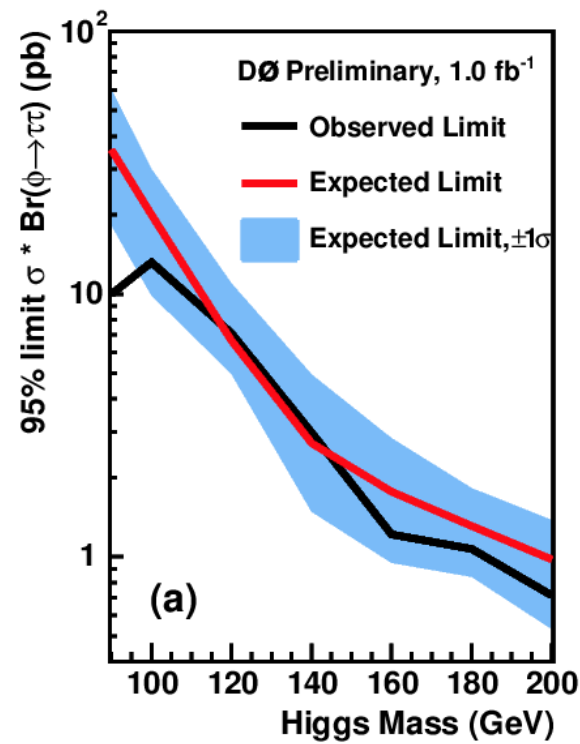}
\caption{The  limit on $\sigma(p\bar{p}\to \phi X)\cdot Br(\phi\to\tau\tau)$ from D0
as a function of Higgs boson mass.}
\label{fig:D0-xsec}       
\end{figure}



 CDF used the binned likelihood fitting technique looking for a peak
in the distribution of the invariant mass of the visible $\tau\tau$
decay products and the vector of the missing energy. A small excess of
events has been observed in the $\tau_e\tau_\mu$ sample at the mass of
about 140~GeV/$c^2$. The statistical significance of the excess is
approximately two standard deviations.


 Neural network analysis has been developed by D0 and used event
parameters such as the visible mass (defined as at CDF), transverse momenta
and pseudorapidities of the decay products. The distribution of the
neutral network output for the data has been compared to the expectation
for the background, and no statistically significant deviation has 
been observed.

\begin{figure*}
\includegraphics[width=0.5\textwidth,angle=0]{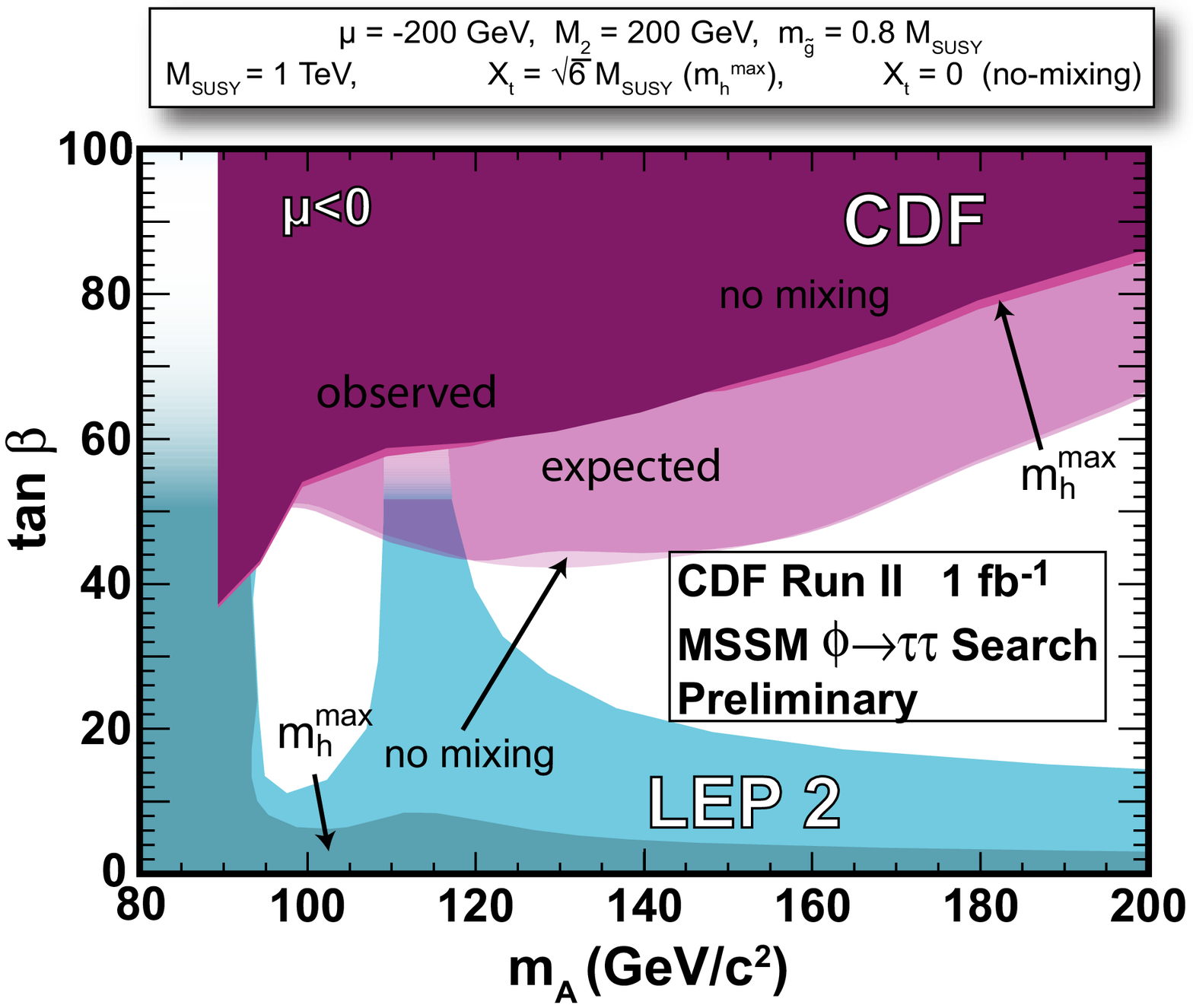}
\includegraphics[width=0.5\textwidth,angle=0]{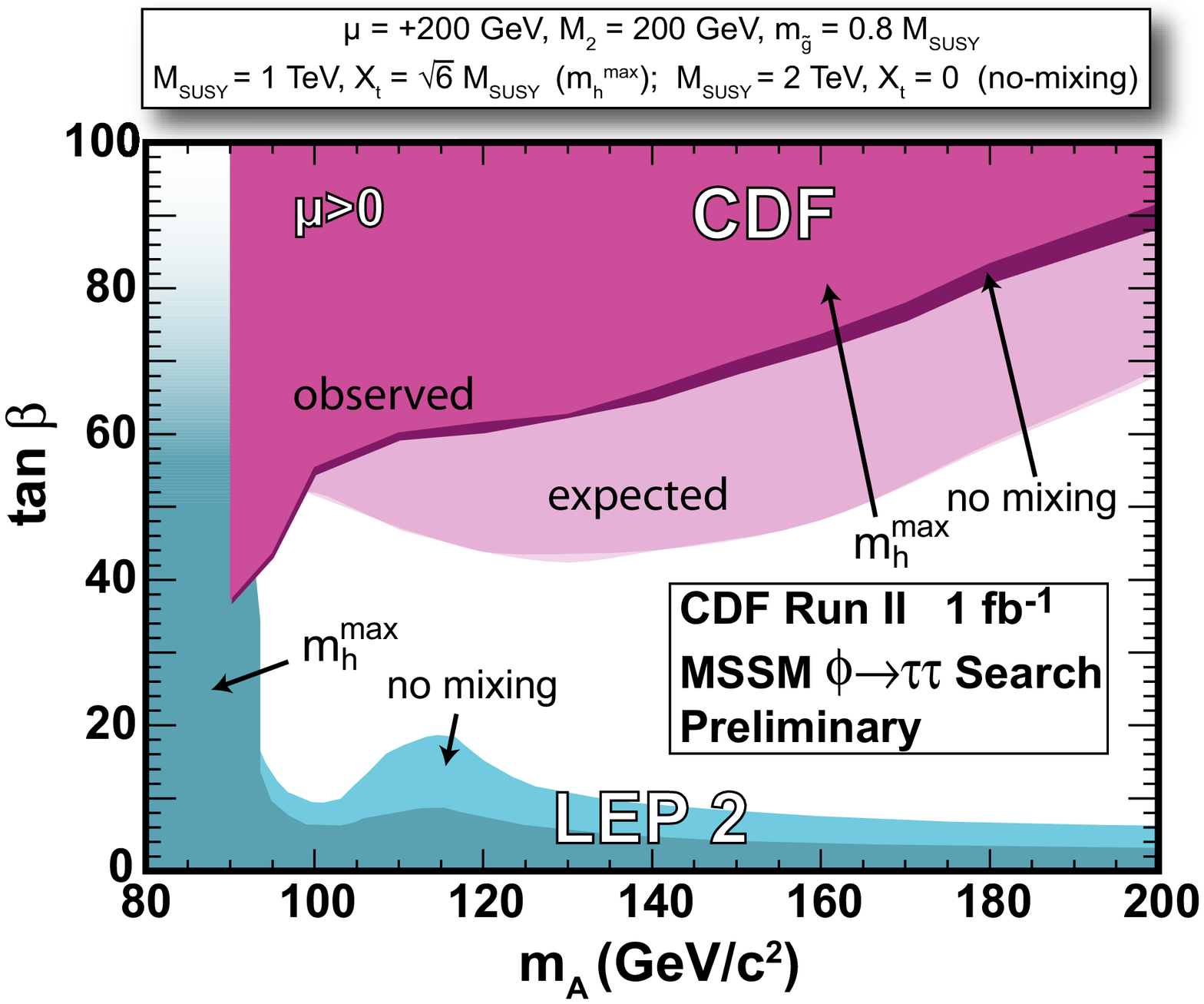}
\caption{CDF exclusion area for $\tan\beta$ as a function of $m_A$ for several benchmark scenarios
  overlaid with the LEP results}
\label{fig:CDF-tanb}       
\end{figure*}

  The data of both CDF and D0 are consistent with background-only
observation. Consequently, both experiments set a 95\%~C.L. exclusion
limit on $\sigma(p\bar{p}\to \phi X)\cdot Br(\phi\to\tau\tau)$, the
product of the Higgs boson production cross-section and the decay rate
of the Higgs to $\tau$-pair. This limit is dependent on the Higgs
boson mass.  The limit from CDF is shown in
Fig.~\ref{fig:CDF-xsec}. The observed exclusion is slightly weaker than
expected for given statistics due to the 2$\sigma$ event excess
mentioned above. The limit from D0 is presented in
Fig.~\ref{fig:D0-xsec}.

\section{Interpretation of Results within MSSM}

  The measured limits on the $\sigma(p\bar{p}\to \phi X)\cdot
Br(\phi\to\tau\tau)$ can be used to derive exclusions on the free
parameters of the MSSM. As the number of the parameters is large, the
common approach to interpreting results of Higgs searches is to fix
most of the MSSM parameters to agreed-upon values and then derive the
exclusion regions in the plane $(m_A,\tan\beta)$, the two parameters
most directly related to MSSM Higgs properties. There are several ways
to fix MSSM parameters corresponding to the most indicative
cases. These are called MSSM benchmark scenarios
\cite{benchmark-scenarios}.  Out of the scenarios described in
\cite{benchmark-scenarios}, CDF and D0 employ the $m_h^{max}$ and the
no-mixing scenario with the case of the mixing parameter of the Higgs
doublets $\mu$ being positive and negative.  The $m_h^{max}$ scenario
is defined as the set of MSSM parameters that maximizes the value of
the mass of the lightest Higgs boson.  The second scenario, no-mixing,
is similar to $m_h^{max}$ , but with the mixing parameter for the stop
quarks set to zero.  Total of four cases are considered.

\begin{figure*}
\includegraphics[width=0.9\textwidth,angle=0]{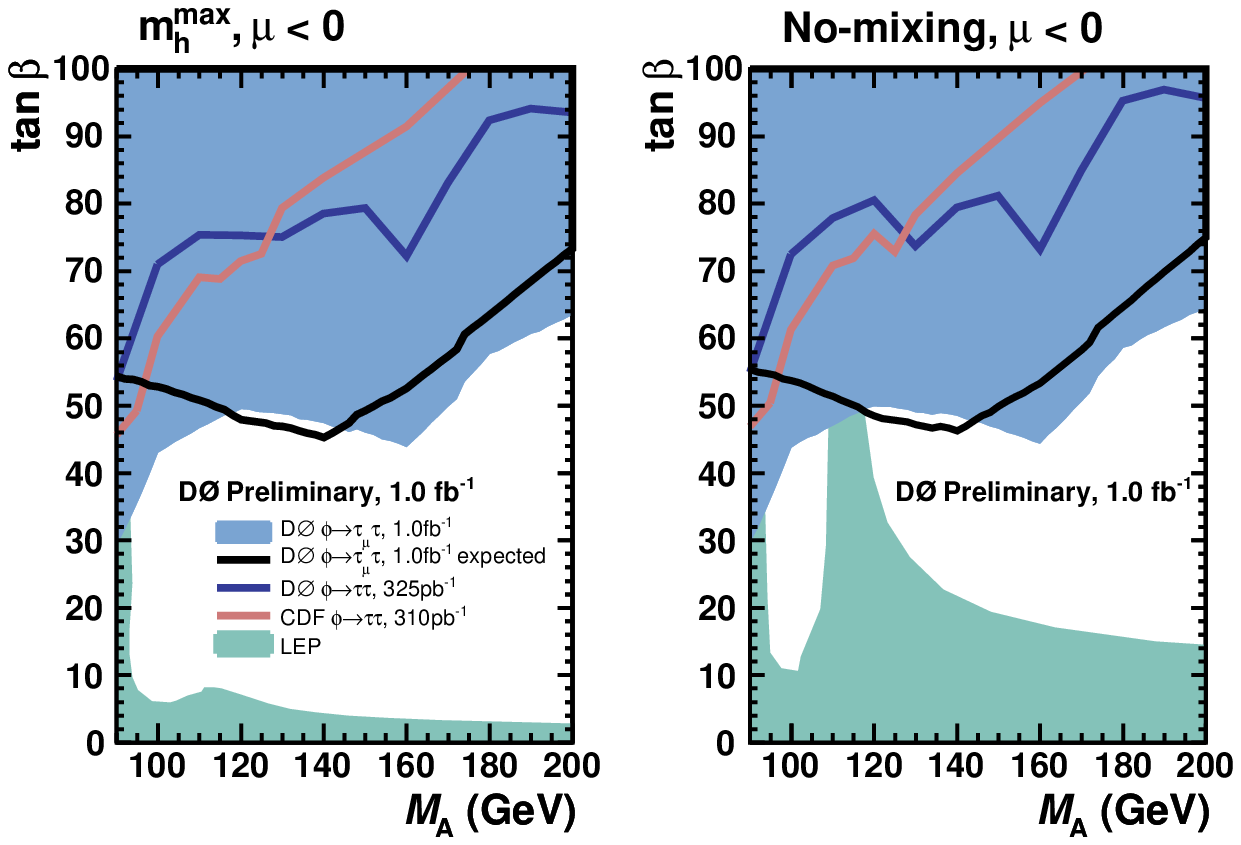}
\caption{D0 exclusion area for $\tan\beta$ as a function of $m_A$ for benchmark scenarios with $\mu<0$
  overlaid with the LEP results and previous MSSM Higgs searches at the Tevatron with lower luminosity
}
\label{fig:D0-tanb-1}       
\end{figure*}

\begin{figure*}
\includegraphics[width=0.9\textwidth,angle=0]{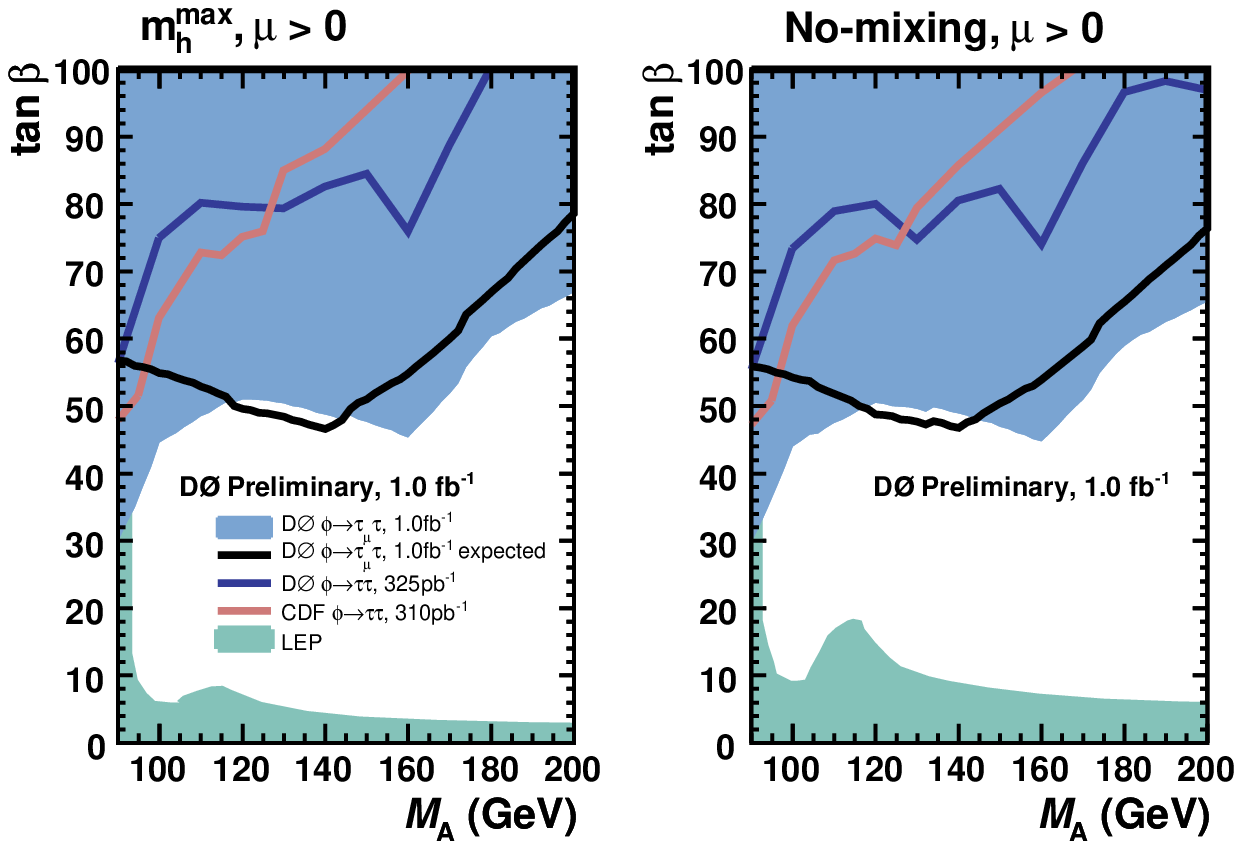}
\caption{D0 exclusion area for $\tan\beta$ as a function of $m_A$ for benchmark scenarios with $\mu>0$
  overlaid with the LEP results  and previous MSSM Higgs searches at the Tevatron with lower luminosity
}
\label{fig:D0-tanb-2}       
\end{figure*}

The CDF exclusion regions for the four scenarios are drawn in
Fig.~\ref{fig:CDF-tanb} and the results from D0 are found in
Fig.~\ref{fig:D0-tanb-1}~and~\ref{fig:D0-tanb-2}. In the studied mass
region the $\tan\beta$ values above 40-60 are excluded in the
considered MSSM scenarios. While no signal is observed and the data
are consistent with the SM backgrounds, the measurements presented
here provide the most constraining to date limits on Higgs decays to
$\tau\tau$ channel.

%
%

\end{document}